%% file: example_paper.tex
\documentclass{article}

\usepackage{microtype}
\usepackage{graphicx}
\usepackage{subfigure}
\usepackage{booktabs} 

\usepackage{hyperref}

\usepackage[accepted]{mlsys2023}

\mlsystitlerunning{FedSS: Federated Learning with Smart Selection of Clients}

\begin{document}

\twocolumn[
\mlsystitle{FedSS: Federated Learning with Smart Selection of Clients}

\mlsyssetsymbol{equal}{*}

\begin{mlsysauthorlist}
\mlsysauthor{Ammar Tahir}{equal,uiuc}
\mlsysauthor{Yongzhou Chen}{equal,uiuc}
\mlsysauthor{Prashanti Nilayam}{uiuc}
\end{mlsysauthorlist}

\mlsysaffiliation{uiuc}{University of Illinois Urbana-Champaign, IL, USA}
\mlsyscorrespondingauthor{Ammar Tahir}{ammart2@illinois.edu}
\mlsyscorrespondingauthor{Yongzhou Chen}{yc28@illinois.edu}

\begin{abstract}
Federated learning provides the ability to learn over heterogeneous user data in a distributed manner while preserving user privacy. However, its current client selection technique is a source of bias as it discriminates against slow clients. For starters, it selects clients that satisfy certain network and system-specific criteria, thus not selecting slow clients. Even when such clients are included in the training process, they either struggle with the training or are dropped altogether for being too slow. Our proposed idea looks to find a sweet spot between fast convergence and heterogeneity by looking at smart client selection and scheduling techniques.
\end{abstract}
]

\printAffiliationsAndNotice{\mlsysEqualContribution}

\input{related}
\input{motivation}
\input{design}
\input{implementation}
\input{evaluation}
\input{future_work}
\newpage

\bibliographystyle{plainnat}
\bibliography{example_paper}

\end{document}

%% file: related.tex
\section{Background}
The past decade has seen an explosion of data-driven and machine learning-based applications that solve different problems. This naturally leads to a lot of fruitful discussions about ownership and access control of these data. Users are concerned about the privacy of their sensitive data and do not prefer sharing it with third-party organizations. However, diverse and heterogeneous training data is of great importance to those models.
Thus, recently we have seen a lot of work on privacy-preserving designs of machine learning models, popularly known as federated learning. 

Federated learning is a privacy-preserving method of distributed learning over heterogeneous user data. Federated learning follows the philosophy of "bringing the code to data, instead of bringing data to code" to address the above-mentioned privacy concerns \cite{fl-scale,mcmahan_ramage_2017}.
The architecture presented by Bonawitz et. al. \cite{fl-scale} consists of a server responsible for selecting client devices for training from a pool of available devices in each round. The server maintains a copy of the global model, which is distributed to selected client devices at the start of each round. Each client trains the model with their local data and sends the gradients back to the server. The server calculates the average gradient using the FedAvg algorithm after collecting gradients. This average gradient is used to update the global model, which is then distributed in the next round.

Since gradients can still reveal information about the data, gradient updates sent by clients are preserved with secure aggregation \cite{bonawitz2016practical} to enforce privacy further. Secure aggregation prevents the server from learning individual clients' gradients but learns the aggregate one once responses from all clients are received. This couples with the already synchronous nature of federated learning, and makes the performance of training susceptible to degradation due to stragglers. Thus, it is desirable from the angle of performance that the selected devices are almost homogeneous with respect to the network and computation power. However, mobile devices can have significantly different network conditions\cite{radiosaber} and compute capability. Most of the current designs select clients randomly from a pool of devices that satisfy certain criteria e.g. minimum 2 GB memory, unmetered network, etc \cite{fl-scale,li2020challenges}. This technique of client selection introduces explicit bias in the system since factors like device memory and quality of network are directly linked with socioeconomic status \cite{abay2020mitigating,kairouz2019advances,li2020challenges}. Thus, it's critical to improve client selection mechanisms to build models that are void of this explicit bias.

\section{Related Work}
A variety of solutions have been proposed for this problem, ranging from model compression to different strategies of client selection. Firstly, we discuss how existing work reduces the training time of slow clients, and still incorporates them to mitigate bias. Then, we discuss current smart client selection strategies.

\subsection{Bias Mitigation Strategies}

As we discussed, the existing design of federated learning introduces explicit bias since parameters based on which devices are selected are linked with socioeconomic factors\cite{abay2020mitigating,kairouz2019advances,li2020challenges}. Bias can also come because of the non-IID distribution of data. However, there are some techniques to circumvent it since it's a deep learning optimization problem\cite{zhao2018federated, li2018federated, power-of-choice}. To mitigate the explicit bias discussed above, we need to include slow devices in the training process. Clearly doing this comes at the cost of slow convergence, owing to the slow clients straggling the process. However, there has been some work on reducing computation and network costs for slow clients at the expense of the accuracy of the model \cite{xu2019elfish,li2018federated,konevcny2016federated}. For example Li et. al. \cite{li2018federated} allows different devices to perform variable amount of workload depending on their resources. Slow clients can run fewer epochs or use lesser input data. Similarly, Kone{\v{c}}n{\`y} et. al. \cite{konevcny2016federated} compresses model updates in a lossy fashion to reduce communication costs at the expense of the accuracy of the sent parameters. Abay et al\cite{abay2020mitigating} proposes a fairness-aware regularization in the loss function. However, we argue that such techniques do not truly address the bias problem because there is still discrimination in how slow clients are handled. An ideal design should give equal opportunity for all clients to contribute to the global model.

\subsection{Clients Selection Strategies}
It's clear that the heterogeneous communication environments and computation resources at clients can hamper the overall training speed. To accelerate the convergence speed under this client heterogeneity, existing work has investigated to make smarter decisions about client selection to alleviate the communication overhead. 

In the prevalent random client selection strategy, all clients participate in the client selection phase in every round. The server uniformly and randomly selects a subset of clients from the pool\cite{communication-efficient,li-partial-selection, flexible-device} for training. The chosen clients do multiple iterations of SGD on the local data, given the ML model and the latest parameters from the server. Finally, the server collects and aggregates the computed gradients to update the global parameters. Li et. al. \cite{li-partial-selection} provides a necessary convergence condition for federated learning on non-iid data with partial client participation. Ruan et. al. \cite{flexible-device} offer a selection scheme that converges even when devices can flexibly join or leave the training.

\begin{figure*}[th!]
    \centering
    \subfigure[Random Selection]{\includegraphics[height=0.21\textwidth]{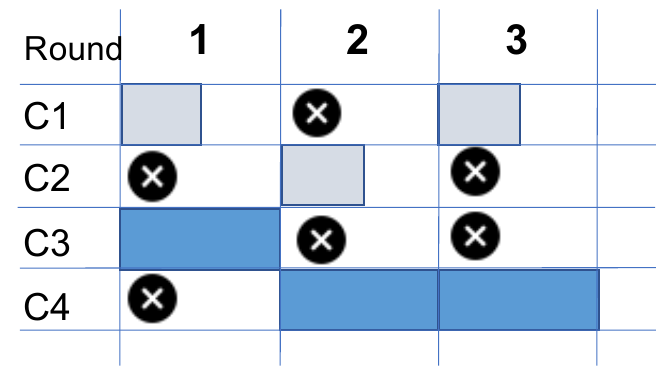}}
    \subfigure[FedCS]{\includegraphics[height=0.21\textwidth]{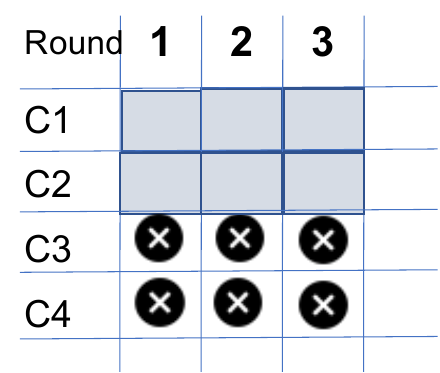}}
    \subfigure[Ours]{\includegraphics[height=0.21\textwidth]{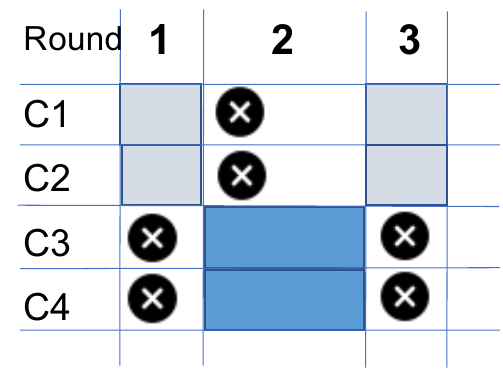}}
    \vspace{-10pt}
    \caption{(a) The round time in the random selection scheme is determined by the straggler. (b) While the round time in FedCS is shortest, client3 and client4 are always excluded because of their long training time and transmission delay. (c) Our scheme reaches the best trade-off between training time and data heterogeneity. }
    \vspace{-10pt}
    \label{fig:motivation}
\end{figure*}

Some recent work looks into client selection based on different criteria like the higher potential for global model convergence or good network conditions. Cho et. al. \cite{power-of-choice} reveal that a biased selection towards clients with higher local loss can increase the speed of convergence. This is because higher loss indicates a higher potential for model improvement. The proposed POWER-OF-CHOICE algorithm can yield up to 3X faster convergence and 10\% higher test accuracy compared with conventional federated learning with random client selection. FedCS\cite{fl-mobile} requests the resource information, such as wireless network bandwidth and compute capability, from selected clients before the distribution of parameters of the global model. It then only collects the gradients from clients which can update and upload the parameters within a deadline. Although these biased client selection models can facilitate quicker convergence, it sacrifices the original benefit of federated learning: the heterogeneity of data. TiFL\cite{chai2020tifl} is another recent work that explores tiering together clients with similar training times and prioritizing faster tiers to speed up training. It only temporarily prioritizes slower tiers when the accuracy of the global model is poor during testing on devices from slower tiers. In contrast, FedSS's client selection does not lean toward any particular cluster during training and offers equal opportunities for every client to contribute to the training.

\subsection{Training Policies}

Recent work also shows that the straggler problem can be eliminated by using asynchronous training policies. For example, in FedAsync \cite{fedasync}, the server does not wait for all clients to send their model updates before performing average. In fact, clients can request a central model whenever they complete their local training. This speeds up training but results in a higher degree of communication and complications due to the staleness of the model \cite{staleness, stripelis2022semi}. Semi-synchronous training \cite{stripelis2022semi} tries to find the best of both worlds. It has a fixed point at which all clients must synchronize at the central server but avoids idling by letting faster clients continue training. However, this is orthogonal to our clustering strategy. Semi-synchronous training can still benefit from our clustering by dividing clients into clusters and having different synchronization deadlines for each cluster. Such clustering will help reduce bias by minimizing extra training that fast clients may do in any round.




%% file: motivation.tex
\section{Motivation}

As evidenced by the related work, federated learning between clients with heterogeneous data, devices, and networks can result in prolonged convergence time. Appropriate client selection decisions can result in quick convergence. The convergence time is determined by two factors: \textit{i) number of rounds} until the model convergence condition is reached and \textit{ii) time duration of each round}. The number of rounds can be reduced by selecting clients that add more value to the learning e.g. high losses. Whereas the duration of each round is determined by stragglers, thus selecting devices based on hardware and network conditions can reduce the convergence time.  

To avoid the prolonged round duration time caused by stragglers, we can select clients with similar training times in each round. Meanwhile, we can randomly choose clients with different network/computation conditions across rounds to guarantee data heterogeneity. We understand that slow clients might be in the minority in some cases, where arranging separate rounds for them can jeopardize their privacy. To ensure privacy for such rounds we can fill up the round group with faster clients too. We visualize and compare this strategy with random selection in Bonawitz et, al. \cite{fl-scale} and FedCS \cite{fl-mobile} in Fig.\ref{fig:motivation}.

%% file: design.tex
\section{System Design}
To reduce bias and also ensure faster model convergence, we propose a smart approach of clients selection. We first collect the device's compute capability captured by FLOPS (Floating point Operations per Second) along with network conditions such as uplink and downlink bandwidth \cite{choi1998priority} of every connected client.

Based on these collected parameters, we categorize clients into different equal sized clusters. Each cluster comprises clients with similar training time. We also determine optimal number of clusters, $k$, for a given distribution. $k$ is optimized to minimize training time, while ensuring higher possible degree of anonymity (high cluster sizes). In every round, the FedSS server chooses a cluster in a round-robin way and selects clients within it randomly to ensure equal opportunity of every client.

There are two basic intuitions behind the approach. 1) Reduce bias due to barriers to entry for low bandwidth/ low-end devices, which in turn minimizes socioeconomic bias. By giving a fair chance to all the clients, the model gets a better chance to learn from different data distributions. This also prevents against content farm attacks using uncompromised phones with high availability and bandwidth. 2) Have more coordinated training rounds with similar performing clients grouped together. We are less likely to run into a situation where the overall completion time of a round is longer due to a fraction of low-performing devices. 


\subsection{Dynamic Client Environment Tracking}
\begin{figure}
    \centering
    \includegraphics[width=0.5\textwidth]{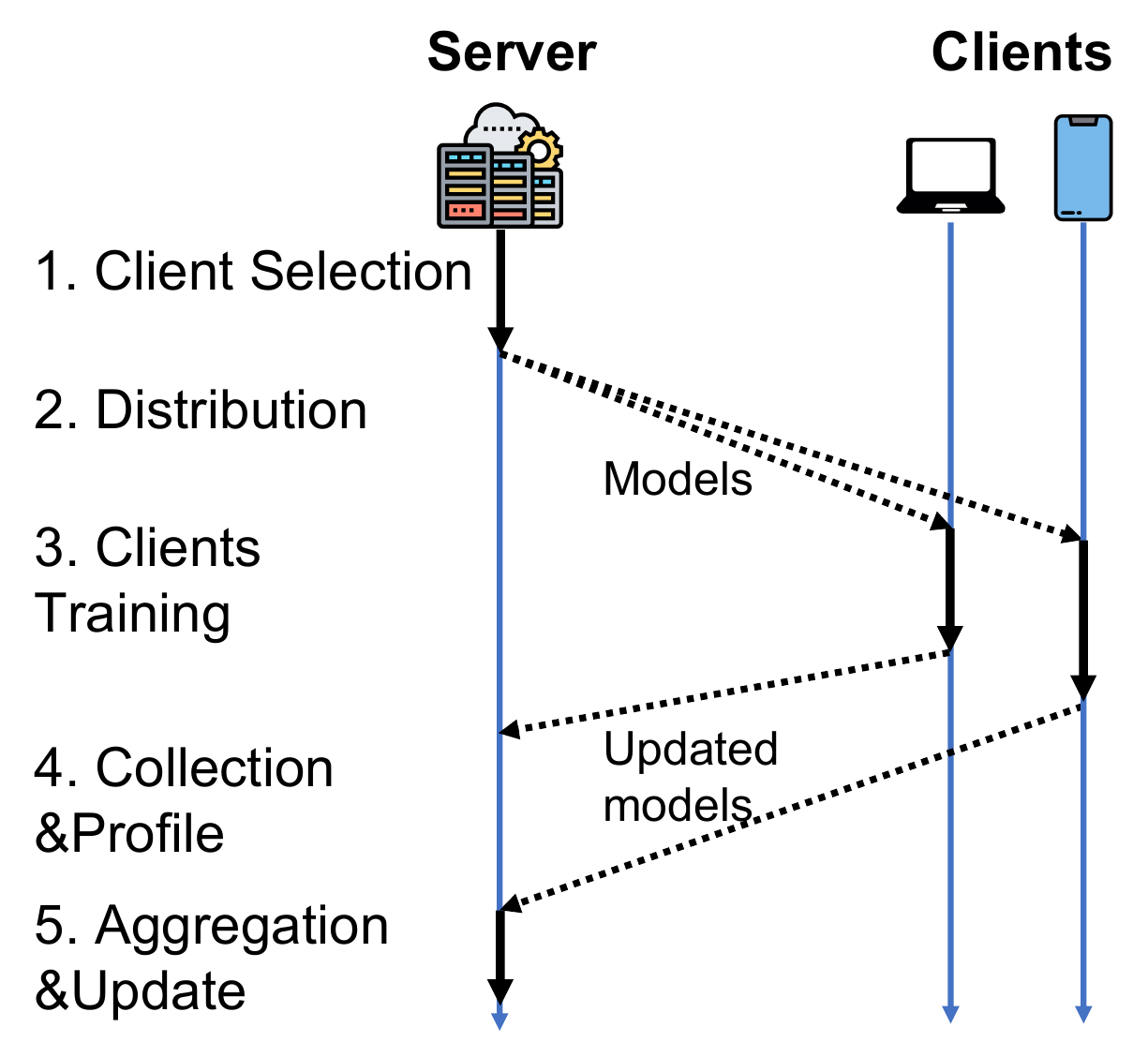}
    \caption{The workflow to finish one round of synchronous training in FedSS. The bottleneck includes model distribution, model collection, and parallel client training, which majorly depend on the client's compute capability and network conditions.}
    \label{fig:overview}
\end{figure}

From Fig.\ref{fig:overview}, the steps in one training round of FedSS comprise: the smart clients' selection, the distribution of models, the parallel clients' training procedure, the collection of updated models, and the aggregation. Except for the first and the last steps, the time it costs in all other steps is decided by the clients' environments. The network conditions, such as bandwidth and propagation delay, determine the time taken to distribute the model and collect gradients. Meanwhile, the client's local compute capability, reflected by available CPUs, GPUs, and memory, determines the time taken to finish the training procedure. Therefore, FedSS dynamically tracks and records those parameters of every connected client, and predicts the approximate time it takes to accomplish one round.

Let's assume the model size is \(M\), and the total number of floating point operations in the model is \(Flops\). Given the measured network uplink bandwidth \(UL_i\), downlink bandwidth \(DL_i\), client's FLOPS rate \(FlopsRate_i\) and the number of samples at client \(S_i\) for client \(c_i\), the overall round time \(T_i\) for that client is:

\[T_i = M / UL_{i} + S_i \cdot Flops / FlopsRate_i + M / DL_{i} \]

There are different faithful methods to measure network bandwidth and FLOPS. Measuring bandwidth has been an active research topic for a long time. There are active ways to measure bandwidth such as speed tests and probe trains \cite{choi1998priority}. There are passive mechanisms as well that estimate bandwidth based on network usage, e.g. client hints \cite{clienthints}, CRAB \cite{tahir2023enabling}. Similarly, there are different benchmarks to measure FLOPS as a proxy of compute capability e.g. Linpack \cite{linpack} and HPL \cite{hpl}. Additionally, these estimates of compute capability can be improved with real-time observations of the time taken to run a training round. The mechanisms to track mentioned metrics are not the contribution of our work, and we rely on existing work for this.



\subsection{Smart Client Selection}
After getting the client round time estimate, $T_i$ for most of the clients, we are ready to run our clustering algorithm. The clustering algorithm takes the number of clusters as an input, and based on it, decides the percentiles of the distribution. For example, if the number of clusters is 3, the selected percentiles would be 25, 50, and 75. Clients are sorted into these clusters based on their $T_i$'s mean-squared distance from percentile values. Since the purpose of this clustering is to group together clients with almost similar training times, equidistant percentile-based centroids serve this purpose reasonably well.

The clusters constructed this way tend to have an uneven distribution of clients. The clusters with a shorter average training time tend to have the most number of clients compared to clusters with a higher average training time. This means if we have a round-robin pattern of switching between clusters for each training round, the clients in larger clusters have a smaller probability of selection. Our strawman solution to this problem was to construct a weighted round-robin scheduling where the weight is based on the cluster size. For example, assume the cluster sizes(out of the total number) are 50\%, 25\%, and 25\% for clusters A, B, and C. The scheduling pattern would be (A, A, B, C), where cluster A is selected for two rounds compared to the others. 

However, we realize that having different-sized clusters also means that clients have varying degrees of privacy. Assuming a malicious federated averaging server, it is easier to deanonymize a participating client in a smaller cluster than a larger one. Thus, another goal of our clustering algorithm is to build clusters of roughly the same size to ensure that the clustering algorithm does not impart any partiality when it comes to privacy. 


We leverage a simple insight to even out clusters created by our algorithm. As we observed that the slow client clusters tend to have fewer clients. We pick the slowest clients from the fast client cluster and migrate them to the slow cluster until sizes are almost evened out. The inverse of this can also be done if the fast client cluster is smaller, but we should be careful so that shifting clients from a slower cluster does not have a very drastic effect.

\label{sec:client-select}

\subsection{Optimal Number of Clusters}

As we pointed out in \ref{sec:client-select}, the standalone client selection algorithm requires the number of clusters as an argument. However, it can be difficult for the operator to know this. Thus, we solve for the optimal value of it. Having fewer clusters is desirable from the perspective of privacy, whereas more clusters result in a shorter average round time. However, the average round time has diminishing returns as we keep increasing the number of clusters. The optimal number of clusters would be the point after which we see diminishing improvement in the average round time. 

\begin{figure}
    \centering
    \includegraphics[width=0.5\textwidth]{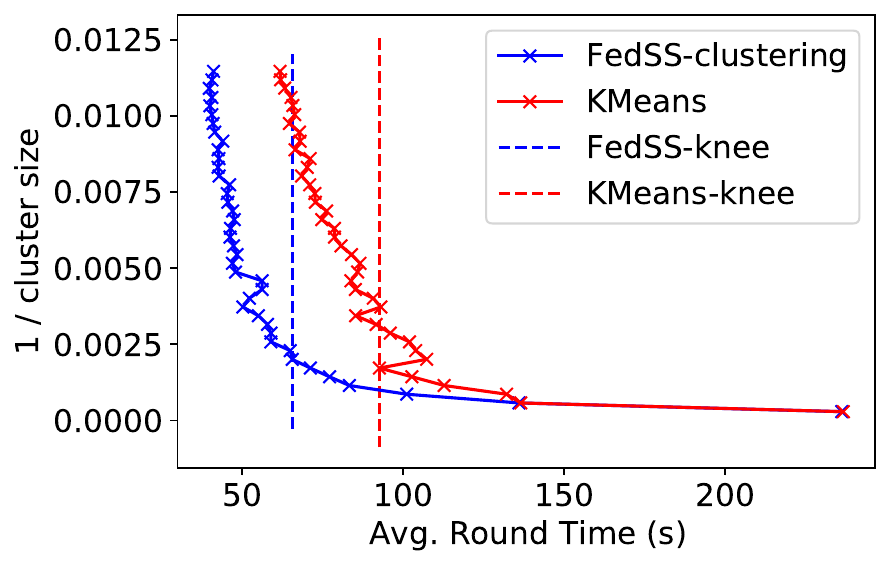}
    \caption{Visualization of the process behind finding the optimal number of clusters (6 in both cases). Also worth noticing is superior efficiency of FedSS's clustering algorithm compared to KMeans.}
    \label{fig:timeVssize}
\end{figure}

Thus, to find the optimal number of clusters, we simulate the training process given round-time estimates of clients with different values of clusters. This gives us the average round time for the given number of clusters. Then, we use Kneedle \cite{satopaa2011finding} to find the optimal point on the curve of average round time vs. 1/cluster size. The Fig.\ref{fig:timeVssize} shows the curve between average round time and 1/cluster size along with the knee point found by the Kneedle algorithm.

Fig.\ref{fig:timeVssize} also compares the effectiveness of our algorithm as compared to KMeans. This is the simulation of training with 10000 clients and 1000 rounds of training. We also compared it against DBScan and KDE clustering, which had almost similar curve as KMeans. The reason our clustering algorithm does better is that it is optimized to reduce round time. Whereas Kmeans and other clustering algorithms are specialized to just cluster data, these clusters may not be optimal with respect to reducing round time. Moreover, it is difficult for clustering algorithms to construct equal-sized clusters. 

\subsection{Overhead}
This grouping algorithm has a time complexity of $NlogN$, where $N$ is the total number of client devices. Finding the optimal number of clusters can be costly if the number of clients is quite high. But this computation can be bounded by limiting the maximum number of clusters to try. For example, it is not desirable to have tens of clusters for only 100 clients. Moreover, the whole algorithm can be run in parallel with model training and a new schedule can be used from the next round onwards. Network conditions as well as training time can vary over time depending on factors like competing traffic, memory, or compute back pressure. Therefore, it is desirable to run the grouping algorithm periodically on updated estimates of $T_i$. 

In our experimental implementation, we did not implement mechanisms to measure bandwidth and profile FLOPS because our experiments are simulation-based. However, existing mechanisms to do so are very lightweight and do not have significant overhead. Similarly, the profiling of FLOPS can be done by measuring the time to train a round. 

%% file: implementation.tex
\section{Implementation}

\begin{figure}
    \centering
    \includegraphics[width=0.5\textwidth]{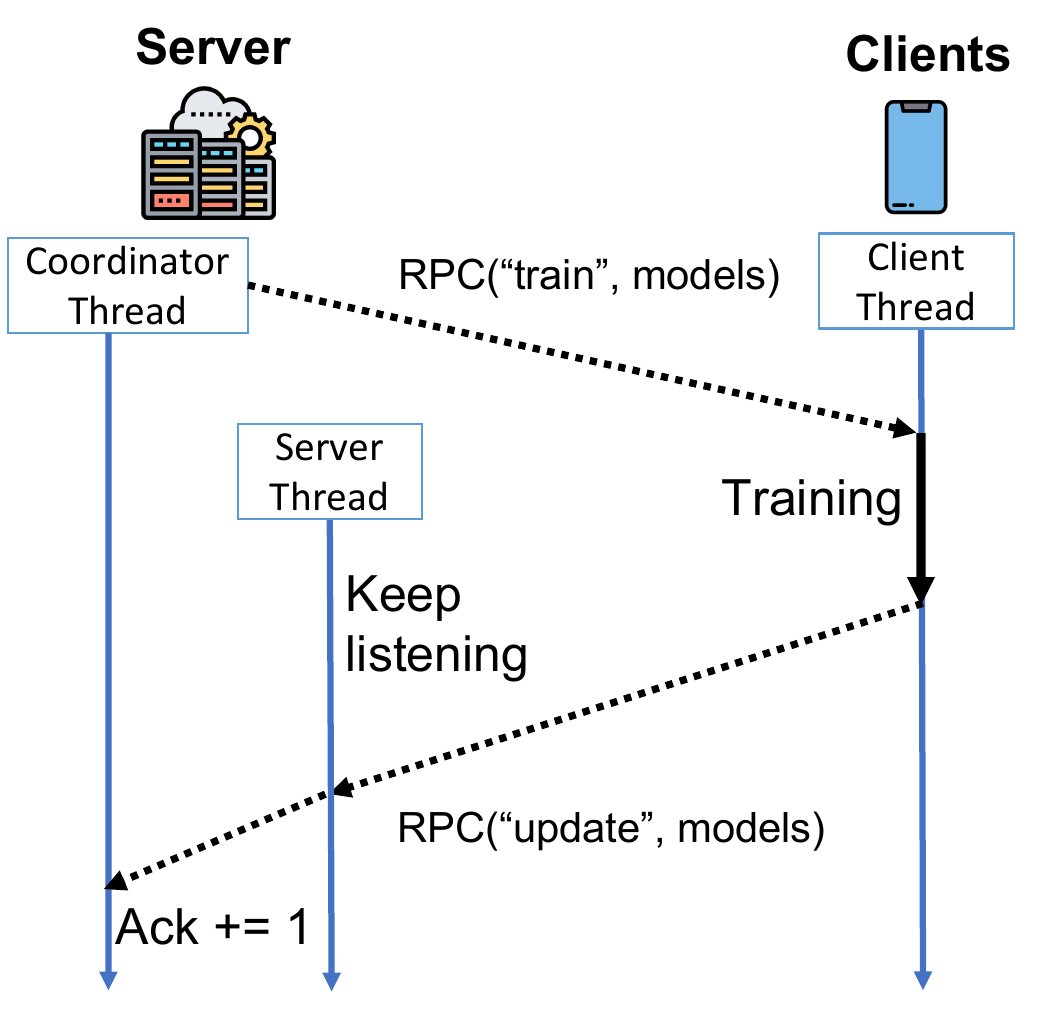}
    \caption{The RPC-based communication protocol}
    \vskip -0.2in
    \label{fig:rpc}
\end{figure}

Based on an open-source federated learning benchmark system leaf\cite{leaf}, we implemented FedSS by writing a RPC-based communication protocol and scheduling logic. We emulate the real federated learning by doing distributed computations in multiple processes on a single machine.

Fig.\ref{fig:rpc} illustrates the communication protocol in detail. The coordinator thread in the server asynchronously sends RPC commands to \(K\) selected clients for training with its local data, while the server thread keeps listening to clients' requests to upload the updated models. After the client finishes training, the server thread receives results and increments a shared variable \textit{acked} by one. The coordinator thread waits until \textit{acked} equals \(K\) to continue the aggregation step and finish this round.

To simulate the variable network, we use Internet Speeds Data from World Population Review \cite{downloadspeeds}. We also simulated the computation time needed for training based on the benchmarked FLOPS \cite{iconcharts} of the top 20 most sold mobile phones in 2020 \cite{yordan_2020}. 



%% file: evaluation.tex
\begin{figure}[t]
    \centering
    \vskip -0.1in
    \includegraphics[width=0.5\textwidth]{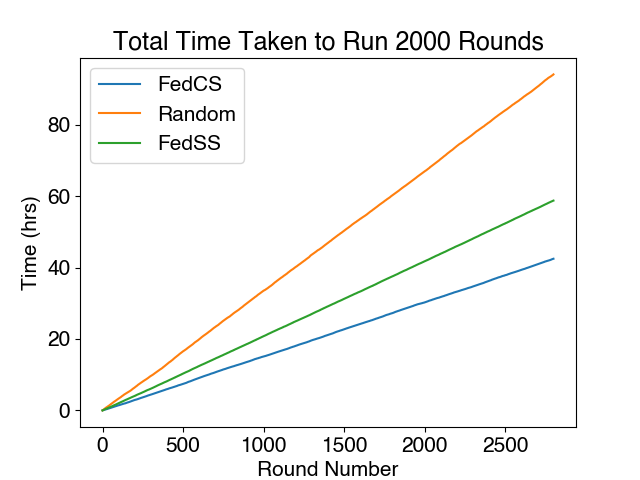}
    \caption{The total time taken by FedSS, FedCS and Random Selection to train 2800 rounds.}
    \vskip -0.2in
    \label{fig:totaltime}
\end{figure}
\begin{figure}[t]
    \centering
    \vskip -0.1in
    \includegraphics[width=0.5\textwidth]{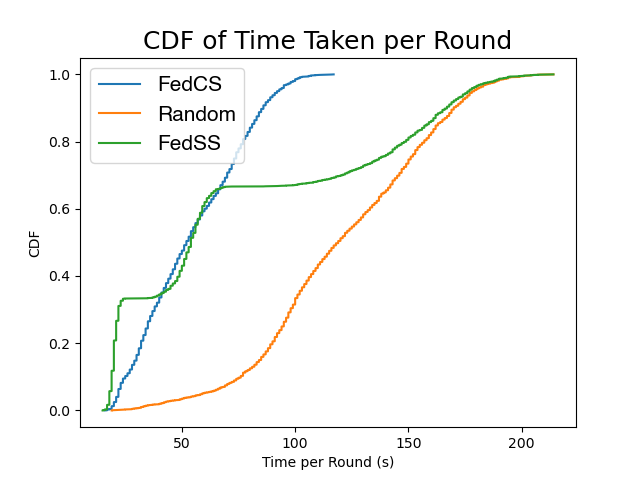}
    \caption{CDF of time taken per round for FedSS, FedCS and Random Selection.}
    \vskip -0.2in
    \label{fig:timecdf}
\end{figure}

\section{Evaluation}

We evaluate FedSS against our implementation of FedCS and random client selection. We train a CNN model on the Femnist dataset, which is distributed in a Non-IID manner across 20 clients. Initially, we aimed to run our experiments with a much larger number of clients, but it's not possible to run so many instances of concurrent CNN models with our limited computation resources. For FedSS and random client selection, we set the number of clients per round to 5. For FedCS, we select 8 clients per round but aggregate updates with the first 5 responses. To cover system and network heterogeneity, we simulate random delays for different clients. However, to ensure fairness between our system and benchmarks, the same delay configurations are used for clients across all our experiments. Similarly, the same distribution of datasets between clients is used across our experiments.

We explored different metrics to evaluate our system. Since our trade-off is between training time and bias, we primarily focus on metrics to capture these two. 

\subsection{Training Time}
Measuring training time is easy, we can individually measure the time taken to complete one round of training at the server. This time would include time taken to send the model to all clients concurrently, and time taken by the clients to train and send back the gradients. This process is bottlenecked by the slowest client accepted by the server for aggregation. Thus, time per each round is measured by the time taken by the slowest client to receive, train and send back the model. 

Fig.\ref{fig:totaltime} shows the total time taken by FedSS and benchmarks to run 2800 rounds of training. FedCS beats our system by approximately 26\% because it only waits for the fastest clients to send back the gradients in every round.
On the other hand, Random Selection has the longest training time, since the time taken per round, in this case, is straggled by the slowest client. FedSS achieves 1.6$\times$ shorter training time than Random Selection. This improvement is possible because of the optimal client clustering according to the training time.

It seems that the training time of all systems is linear to the number of rounds in Fig.\ref{fig:totaltime}. This is due to too many data points and doesn't indicate that the training time of every round is similar. Fig.\ref{fig:timecdf} shows the CDF of training time per round for all systems. Steps in the line for FedSS show different training times for different clusters (3 clusters in this particular case). FedSS, which selects clients with similar training time together, is able to finish 80\% of the rounds faster than Random and save around 40 hours.

\begin{figure}[t]
    \centering
    \includegraphics[width=0.5\textwidth]{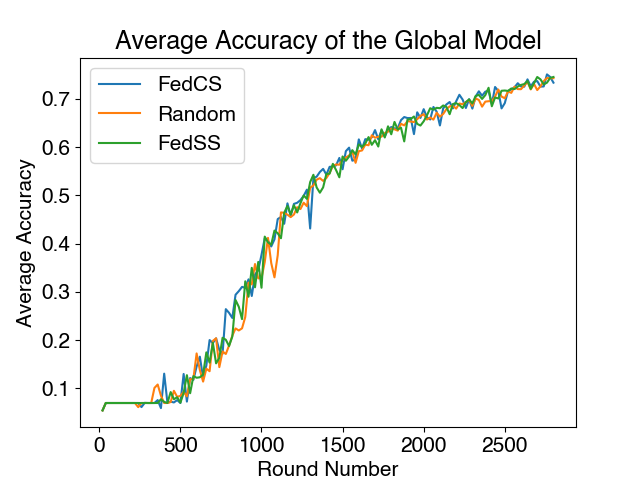}
    \caption{The average accuracy of the global model for FedSS, FedCS and Random Selection.}
    \label{fig:accuracy}
\end{figure}
\begin{figure}[t]
    \centering
    \includegraphics[width=0.5\textwidth]{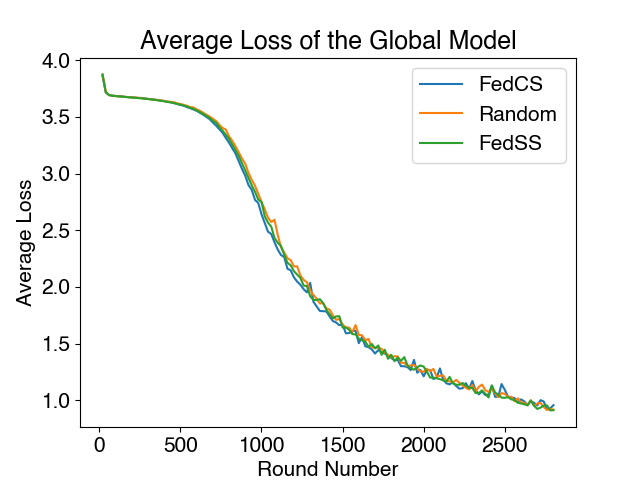}
    \caption{The average loss of the global model for FedSS, FedCS and Random Selection.}
    \label{fig:loss}
\end{figure}

\subsection{Bias}
Before measuring bias, let's take a look at the performance of the global model after 2800 rounds for each of the selection strategies. There are many well-established evaluation metrics for classification models including precision, recall, accuracy, F1-score, etc. For our evaluation, we select accuracy and F1-score as the metrics.

\subsubsection{Accuracy}
Accuracy is the metric to determine the correct prediction ratio for the given dataset. It is calculated as :
\[\frac{TP+TN}{TP+TN+FP+FN}\]

Fig.\ref{fig:accuracy} \& \ref{fig:loss} respectively show the average accuracy and loss of the global model across all clients. From the first look, it seems as if all the strategies have similar performance with respect to model validity. However, a deeper look at the model's metrics shows a clearer picture. 

\begin{table*}
\centering
\begin{tabular}{|c|ccc|c|ccc|}
\hline
\multicolumn{4}{|c|}{Accuracy for 4 Slowest Clients}                                                                            & \multicolumn{4}{c|}{Accuracy for 4 Fastest Clients}                                                                             \\ \hline
Client  & \multicolumn{1}{c|}{FedCS} & \multicolumn{1}{c|}{\begin{tabular}[c]{@{}c@{}}Random \\ Selection\end{tabular}} & FedSS & Client  & \multicolumn{1}{c|}{FedCS} & \multicolumn{1}{c|}{\begin{tabular}[c]{@{}c@{}}Random \\ Selection\end{tabular}} & FedSS \\ \hline
1  & 68.76  & \textbf{80.12}  & 79.81 & 1  & 83.28   & 85.92     & \textbf{86.80} \\
2  & 72.34  & 76.17 & \textbf{76.59} & 2  & 55.97   & 55.97     & \textbf{60.37} \\
3  & 56.68  & 67.51  & \textbf{71.33} & 3  & \textbf{80.07}   & 76.56     & 78.90 \\
4  & 67.32  & \textbf{71.25}  & 70.86 & 4  & \textbf{84.88}   & 79.42    & 78.13 \\ \hline
Average & 66.27     & 73.76    & \textbf{74.65} & Average & \textbf{76.05}    & 74.47    & \textbf{76.05} \\ \hline
\end{tabular}
\caption{A breakdown of the accuracy of the model across 4 slowest and 4 fastest clients for each of the selection strategies.}
\label{table:bias-accuracy}
\end{table*}

Table\ref{table:bias-accuracy} shows the breakdown of the accuracy of the model across different kinds of clients. We notice that while FedCS has an average accuracy as good as Random Selection and FedSS, it has lower accuracy for slower clients. On the other hand, its accuracy for faster clients is similar to both Random Selection and FedSS. This difference in performance on slow clients captures the bias in FedCS's client selection strategy. Fast clients end up getting a lot more training opportunities than slow clients.

The 1.5\% accuracy difference between FedSS and Random Selection, although negligible, can be explained from a data heterogeneity point of view. Recall that Random Selection by design does not have any explicit bias due to device heterogeneity. This is because Random Selection selects clients randomly and then waits for all of them to respond before aggregating. Thus, the 1\% difference in its accuracy when compared to faster vs slower clients, can solely be attributed to data heterogeneity. The fact that FedSS does as good as Random Selection while reducing the training time by a huge margin shows promise that FedSS can reduce bias and amortize the cost of slow nodes.

\subsubsection{F1 Score}

F1-score is a metric to determine model performance by using both, precision and recall values. Precision is the metric to determine the percentage of correct results out of all the results for a given class. It is given as : \[ TP/ (TP+FP) \]  
Recall is the metric to determine the percentage of correct results out of all the true results for a given class. It is given as : \[ TP/ (TP+FN) \]
F1 score, also known as the balanced F score, is the harmonic mean of precision and recall.
\[2*\frac{precision*recall}{precision+recall}\]
For a multi-class classification, the F1 score is given as the average of the F1 score of each class based on averaging scheme. For our experiments, we have selected \textbf{weighted} averaging to deal with class imbalance issues due to the non-iid distribution of data across multiple clients.

\begin{table*}
\centering
\begin{tabular}{|c|ccc|c|ccc|}
\hline
\multicolumn{4}{|c|}{F1 score for 4 Slowest Clients}                                                                            & \multicolumn{4}{c|}{F1 score for 4 Fastest Clients}                                                                             \\ \hline
Client  & \multicolumn{1}{c|}{FedCS} & \multicolumn{1}{c|}{\begin{tabular}[c]{@{}c@{}}Random \\ Selection\end{tabular}} & FedSS & Client  & \multicolumn{1}{c|}{FedCS} & \multicolumn{1}{c|}{\begin{tabular}[c]{@{}c@{}}Random \\ Selection\end{tabular}} & FedSS \\ \hline
1  & 67.10 & \textbf{78.08} & 77.22 & 1 & 81.48   & 83.72     & \textbf{84.45} \\
2  & 69.00 & 72.58 & \textbf{73.51} & 2 & 56.34   & 56.86    & \textbf{60.85} \\
3  & 54.04 & 66.32 & \textbf{69.97} & 3 & \textbf{77.50}   & 73.60    & 75.64 \\
4  & 62.28 & \textbf{68.02} & 67.69 & 4 & \textbf{85.13}   & 79.39    & 78.14 \\ \hline
Average & 62.60  & 71.25 & \textbf{72.09} & Average & \textbf{75.11}    & 73.39 & 74.77 \\ \hline
\end{tabular}
\caption{A breakdown of F1 score of model across 4 slowest and 4 fastest clients for each of the selection strategies.}
\label{table:f1}
\end{table*}

\begin{figure}[t]
    \centering
    \includegraphics[width=0.5\textwidth]{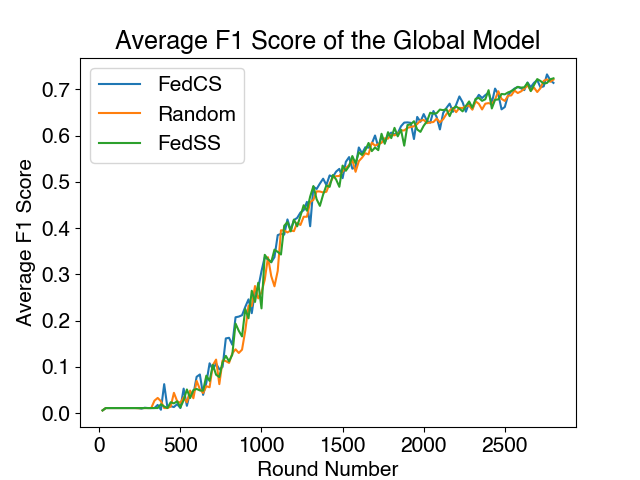}
    \caption{The average F1 score of the global model for FedSS, FedCS and Random Selection.}
    \label{fig:f1score}
\end{figure}

Table\ref{table:f1} shows the breakdown of the F1-score for the slowest and the fastest clients trained in our experiment. The results are consistent with the accuracy scores. FedSS achieves results similar to FedCS for the fastest clients and improves the results by 16\% for the slowest clients as compared to FedCS. Similar to accuracy, the results from FedSS match Random Selection while taking considerably less amount of time to train.

%% file: future_work.tex
\section{Future work}


We discuss the limitations of FedSS and the future work in this section. While FedSS achieves the best trade-off between training time and data heterogeneity in our testbed, we have some restrictions in experiments that need more work and real-world experimentation.

\textbf{Experimental Setup: }
Currently, we are bound to work on our local setup, which restricts us to be able to run experiments with only 20 clients. Running with more clients results in severe resource congestion(GPUs and memory). We expect better performance with more clients with our smart clustering and selection. Also, it would be really valuable to be able to run our setup with some real devices, along with implementations to track the client environment. 

\textbf{Handling Churn: } Currently FedSS assumes that all clients join in the beginning and maintain available during the whole training procedure, which may not be true in real-world federated learning. Recalling that it takes a few rounds for FedSS to measure and profile the new client's network condition and compute capability, it can be hard for FedSS to track this information accurately in the scenario where clients join and leave frequently.

To mitigate the possible performance degradation due to the measurement and profiling delay of newly connected clients, we propose to collect the hardware specifications of clients, such as processor types and memory size, to initialize the grouping based on some regression models. The rationale behind it is that clients with similar hardware should have similar computation capability under most circumstances (unless many other background tasks compete for resources). Whenever a new client joins, FedSS takes one RTT to collect this information and determine its initial group based on the model, which is still better than random initialization.

\textbf{Data Heterogeneity: } FedSS treats all clients equally in every selection, which may not be the optimal solution. Although the selection is unbiased, the data heterogeneity itself among clients is another important source of bias\cite{kairouz2019advances, zhao2018federated, jeong2018communication}. If a subset of clients with similar kinds of data is picked most of the time, the model will converge faster but can have a severe bias to the data distribution of that subset and may not be able to capture the true global data distribution.

To overcome this issue, we want to explore the possibility to introduce the training loss of individual clients into the client selection process. Specifically in every round, if the training loss for a particular client is non-significant, we exclude it from selection for the next few rounds. This will improve the probability of other clients contributing towards the global model and also prevent model askew.

\textbf{Scalability and Fault-tolerance} FedSS applies only one server in the system, which induces a communication bottleneck and a single point of failure. We plan to leverage locality-aware multi-server architecture to enhance scalability, fault-tolerance, and even training performance in the future.

Firstly, a locality-aware server tends to schedule and disseminate its model to local clients with better network service, which alleviates the communication bottleneck on the server side and reduces the ratio of stragglers resulting from the network. Secondly, multiple servers can apply a consensus algorithm to replicate the training state of each other to tolerate failure. Lastly, for geographically distributed applications such as next work prediction\cite{hard2018federated} and geo-local language translation\cite{nord2005text}, a locality-aware server can capture and preserve those geo-local characteristics better.


\section{Conclusion}

Federated learning due to its distributed nature of training machine learning model suffers from the issues of data and device heterogeneity. When we talk about device heterogeneity, there exists a trade-off between short training time and bias, as existing schemes end up dropping slower clients. We show this trade-off by comparing existing mechanisms of client selection. We then argue that to eliminate bias, it is necessary to make slow clients part of training. We present FedSS which finds a sweet spot between short training time and handling device heterogeneity by performing a smart selection of clients. 